# A METHOD TO SPEED UP CONVERGENCE OF ITERATIVE LEARNING CONTROL FOR HIGH PRECISION REPETITIVE MOTIONS

Richard W. Longman,[1] Shuo Liu,[2] and Tarek A. Elsharhawy[3]

Various spacecraft have sensors that repeatedly perform a prescribed scanning maneuver, and one may want high precision. Iterative Learning Control (ILC) records previous run tracking error, adjusts the next run command, aiming for zero tracking error in the real world, not our model of the world. In response to a command, feedback control systems perform a convolution sum over all commands given since time zero, with a weighting factor getting smaller going further back in time. ILC learns to eliminate this error through iterations in hardware. It aims to find that command that causes the real world system to actually follow the desired command. The topic of this paper considers the possibility of learning to make our model of the world produce less error. This can be done easily and quickly numerically, and the result then used as a starting point for the ILC iterations performed in hardware. The point is to reduce the number of hardware iteration, converging more quickly to closely tracking the desired trajectory in the world. How to decide how many iterations with our model to make before switching to hardware iterations is presented, with numerical simulations performed to illustrate the approach.

## INTRODUCTION

Spacecraft often have sensors whose task is to make repeated scans, and one may want the scans to be made with high tracking precision. Classical control systems are imperfect, in mathematical terms, given a time varying command, the response is a convolution integral over all previous commands. The sum has a weighting factor that decays with time like the time constants in the system. This factor is a pure weighting factor for systems with all real roots to the associated characteristic polynomial, but gets more complicated when there are resonances involved. The feedback response error is the difference between the current command which is associated with the upper limit of this integral, and the integral over all this and all previous commands. Iterative Learning Control (ILC) aims to learn how to eliminate this error and make the control system actually perform the command it is given (Refs. 1-5).

ILC learns from experience observing the error associated with a command, and then adjusts the command for the next run or iteration, aiming to converge to that input that actually produces the desired output. The approach can be very effective. Experiments performed on the Robotics Research Corporation robot at NASA Langley shown in Fig. 1, reduced the tracking error by about 3 orders of magnitude in 12 iterations for learning, when performing a high-speed maneuver of all 7 joints (Ref. 6). This example has particularly fast learning. Other systems can require more iterations, and the point of this paper is to present a method that can speed up the learning in such cases.

[1] Professor, Mechanical Engineering Department, and Civil Engineering and Engineering Mechanics Department, Columbia University, 500 West 120th St., New York NY 10027, USA
[2] Doctoral Student, Department of Mechanical Engineering, Boston University, 110 Cummington Mall, MA 02215 USA
[3] Lecturer, Cal Poly Pomona, College of Engineering, 3801 W. Temple Ave., Pomona, CA 91768 USA

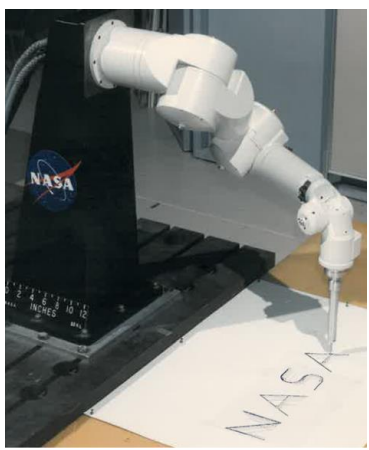

Figure 1. Robotics Research Corporation Robot

ILC is an inverse problem, given the desired output, find the input needed to produce it, and do so in the real world by iterating with the real world in repeated runs. The initial run can be any input, and one observes the associated output. From any such starting point, ILC iterations are to progressively change the input run by run, each time reducing the tracking error. There are various possible ILC laws to accomplish this, including what is here called the *P* Transpose law (Ref. 7), the Partial Isometry law (Ref. 8), and the Quadratic Cost learning law referred to as the Norm Optimal law (Ref. 9). All of these laws can be put under one mathematical umbrella as shown in (Ref. 10). Each law is designed to be numerically robust to model error, e.g. the *P* Transpose law makes an update matrix that is numerically symmetric, producing orthonormal real eigenvectors. Robustness to model errors can be further improved by reducing the learning gain, and this increases the number of iterations needed for convergence. During the ILC iterations one may need to reduce the ILC learning gain in order to obtain a convergent process.

The discrete time inverse problem has some issues that are often not realized by users. For discrete time systems that are the result of feeding a continuous time system through a zero order hold, the result can often require an unstable control action to produce the desired output (Ref. 11). This may go unnoticed because the instability can grow slowly and people stop the iterative process before seeing it (Ref. 12). The result is perhaps substantial error improvement but still a disappointing final error level. However, we have methods to bypass this difficulty by not asking for zero error in the first or first few time steps (Refs. 14-16).

Usually, our model of the world is reasonably accurate in some range of frequencies. Consider that we apply ILC iterations to our model of the world. The model predicted error will be reduced, and the learned command applied to the world will have error improvement in frequency ranges where our model is relatively accurate. The ILC iterations with our model can be done purely numerically on a computer and can even be accelerated easily. They do not require any experiments to be performed in hardware. An acceleration method was first introduced in Ref. (18) and the speedup concept of this paper says: make ILC iterations using our model of the world to quickly produce a command input with substantially reduced error, then apply the input obtained as the initial run to start ILC iterations with the real world. The point is that many time-consuming experiments applying current candidate ILC updates can be bypassed. A formula is developed so that one can directly compute the result of a chosen number of iterations with our model, and there is then no need to actually perform such iterations on our model to get the improved starting point for the hardware iterations with the world.

There remains the issue of how to know how many iterations to consider with the model before applying the resulting command input to the world starting iterations with the world hardware. A criterion is presented to help make this decision. Numerical examples are presented to illustrate the process and the resulting behavior.

## FINITE TIME INPUT-OUTPUT MODEL

Iterative Learning Control is a finite time problem that aims to produce zero tracking error throughout the initial transients as well as later to the end of the trajectory. Here we use the matrix approach packaging all time steps of an ILC iteration into column vectors, as in Ref. (15). Consider a single-input single-output system in state variable form

$$x(k+1) = Ax(k) + Bu(k) \quad k = 0,1,2,\ldots,N-1 \quad (1)$$
$$y(k) = Cx(k) \quad k = 1,2,3,\ldots,N$$

where $\underline{u}_j = [u_j(0), u_j(1), \ldots, u_j(N-1)]^T$ is the input history vector at iteration or run $j$, and $\underline{y}_j = [y_j(1), y_j(2), \ldots, y_j(N)]^T$ is the resulting output history vector from that run. The desired output and the associated output error are denoted by

$$\underline{y}^* = [y^*(1), y^*(2), \ldots, y^*(N)]^T \quad (2)$$
$$\underline{e}_j = \underline{y}^* - \underline{y}_j$$
$$\underline{e}_j = [e_j(1), e_j(2), \ldots, e_j(N)]^T$$

Writing the convolution sum solution of Eq. (1) for all time steps and packaging it in matrix form

$$y_j(k) = CA^k x(0) + \sum_{i=0}^{N-1} CA^{k-i-1} Bu_j(i) \quad (3)$$

$$P = \begin{bmatrix} CB & 0 & \cdots & 0 \\ CAB & CB & \cdots & \vdots \\ \vdots & \vdots & \ddots & 0 \\ CA^{N-1}B & CA^{N-2}B & \cdots & CB \end{bmatrix} ; \quad \bar{A} = \begin{bmatrix} CA \\ CA^2 \\ \vdots \\ CA^N \end{bmatrix}$$

results in the finite time input-output relationship used here

$$\underline{y}_j = P\underline{u}_j + \bar{A}x(0) \quad (4)$$

## ILC IS AN INVERSE PROBLEM

The aim of ILC is: Given $\underline{y}^*$ find $\underline{u}^*$ such that

$$\underline{y}^* = P\underline{u}^* + \bar{A}x(0) \quad (5)$$

i.e., given the desired output, find the input that will produce it. Start from an initial run with a chosen input history $\underline{u}_0$ which can be any function such as zero or the desired output. Apply it to the hardware and observe the resulting output $\underline{y}_0$. Based on the error $\underline{e}_0$ observed, modify the input according to an ILC law for each succeeding run

$$\underline{u}_{j+1} = \underline{u}_j + L\underline{e}_j \tag{6}$$

where $L$ is an ILC learning gain matrix. Subtract Eq.(4) from $\underline{y}^*$, making the left hand side equal to $\underline{e}_{j+1}$ and use Eq. (6) to produce the error propagation equation from iteration to iteration

$$\underline{e}_j = (I - PL)\underline{e}_{j-1} \tag{7}$$

The error will converge to zero as $j \to \infty$ for all possible $\underline{e}_0$ if and only if the spectral radius of matrix $(I - PL)$ is less than unity, i.e. is $|\lambda_i(I - PL)| < 1$ for all eigenvalues. It will converge monotonically in the sense of the Euclidean norm of the error vector if the maximum singular value of this matrix is less than unity, $\sigma_i(I - PL) < 1$ for all $i$. Equation (7) gives the error at each $j$, and inserting it into Eq. (6) gives the input that produced this error. This produces the history of all inputs and all outputs for each $j$ as the iterations progresses.

There are many choices for the ILC law. The early versions by Arimoto simply adjusted the input by a gain times the time step one period back but shifted one step forward to account for the one step time delay through the system. This is his $P$ type ILC (Ref. 1). He also gave a $D$ type. Both have the wonderful property that they are model independent. But there is a big price to pay for this, it likely results in bad transients during the learning process. More recent ILC laws make use of one's model of the system as given in matrix the $P$ of Eq. (4), using some knowledge of the system to obtain well behave transients during the learning iterations. Following are three effective ILC laws

$$L = \phi P^T \tag{8}$$

$$L = \phi V U^T \ : \quad P = U \Sigma V^T \tag{9}$$

$$L = (\phi I + P^T P)^{-1} P^T \tag{10}$$

The first of these we refer to as the $P$ Transpose ILC law (Ref. 7), the second of these is the Partial Isometry law (Ref. 8), and the third is a simple case of the Norm Optimal ILC law (Ref. 9). The $\phi$ in each law is a learning gain chosen by the user to ensure convergence. All three laws can be viewed as special cases of the Norm Optimal law as demonstrated in (Ref. 10). Note that all three of these laws produce a symmetric matrix $(I - PL)$, giving good numerical robustness properties. All eigenvalues are real and positive, and all eigenvectors are real and orthogonal.

**THE INITIAL RUN**

In the initial run one applies an arbitrary input $\underline{u}_0$ to the hardware, and observes the resulting output $\underline{y}_0$ with the corresponding error $\underline{e}_0 = \underline{y}^* - \underline{y}_0$. One can pick the input as all zero, which has some advantages, or if matrix $P$ represents a feedback control system, one might prefer to make $\underline{u}_0$ be the desired output $\underline{y}^*$. Feedback control systems do not do what you ask. The concept of bandwidth finds what error is made after reaching steady state response, when commands are sinusoids. We can get a more nuanced assessment of what tracking errors feedback control systems make by examining the following simple problem (done in the continuous time domain, but the properties also hold for discrete time systems). Consider the following first order plant equation

$$\frac{dy}{dt} + ay = u \tag{11}$$

with a proportional controller

$$u = k(y^* - y) \tag{12}$$

producing the following feedback control model,

$$\frac{dy}{dt} = -(a+k)y + ky^* \tag{13}$$

Note that this is written in the form of a one dimensional state space model. The solution to a state space model equation is

$$y(t) = e^{-(a+k)t}y(0) + \int_0^t e^{-(a+k)(t-\tau)}ky^*(\tau)d\tau \tag{14}$$

which can be rewritten with a change of variable $\tau' = t - \tau$ as

$$y(t) = e^{-(a+k)t}y(0) + \int_0^t e^{-(a+k)\tau'}ky^*(t-\tau')d\tau' \tag{15}$$

The first term is the solution of the homogeneous equation and is unrelated to the command. The second term is the particular solution response to command $y^*$. The actual output at time $t$ is the left hand side of the equation. The desired output is $y^*(t)$ which is only seen in the particular solution at the upper limit of the integral. The particular solution produced is then an integral over all commands $y^*(\tau)$ from time zero to the present time, and one certainly cannot expect the integral to be equal to $y^*(t)$ at the endpoint. Note that the integral is the solution with zero initial conditions, so there are transients included in the integral. The $e^{-(a+k)\tau'}$ can be thought of as a weighting factor, with greatest weight given to the most recent time, and the weight decays going backward in time according to the time constant of the exponential. One can generalize this, for higher order differential equations with real roots, producing multiple integrals in succession each with its own weighting factor and its own time constant. To generalize to systems with complex roots, the behavior is worse compared to the desired output, as the weight factor can fail to be positive resulting in resonance with past inputs.

It is this error that ILC aims to eliminate, by replacing the command by some new function which makes the integral together with the initial condition term actually equal $y^*(t)$ for all time steps in the finite time, discrete time problem.

**THE USUAL IMPLEMENTATION**

Implementation requires the ILC law to be based on our knowledge of the system

$$\underline{u}_{j+1} = \underline{u}_j + L_M \underline{e}_j \tag{16}$$

The subscript $M$ on the learning gain matrix $L_M$ indicates that our model of the world denoted by $P_M$ is used in whichever ILC law we decide to use. But the ILC update is made applying the update to the world whose dynamics is given by $P_W$ which we do not know. The ILC Laws become

$$L_M = \phi P_M^T \tag{17}$$

$$L_M = \phi V_M U_M^T \quad : \quad P_M = U_M \Sigma_M V_M^T \tag{18}$$

$$L_M = (\phi I + P_M^T P_M)^{-1} P_M^T \tag{19}$$

Convergence to zero error for all initial $\underline{u}_0$ and $\underline{y}_0, \underline{e}_0$ obeys

$$\underline{e}_j = (I - P_W L_M)\underline{e}_{j-1}$$

and is guaranteed to converge to zero error for all initial run values, provided $|\lambda_i(I - P_W L_M)| < 1$, and to do so monotonically as seen in the Euclidean norm if $\sigma_i(I - P_W L_M) < 1$, for all $i$.

Because we do not know $P_W$, the ILC designer must pick the learning gain $\phi$ wisely. Since without model error the ILC laws are robust numerically, with a sufficiently small gain, one expects convergence to zero error in the presence of model error. There are two categories of error: Error in the coefficients of one's model and missing high frequency modes. The expectation is very likely right for the first category, and the designer simply keeps the gain sufficiently small that error in the hardware runs with the world is decreasing. In the second case, high frequency modes missing from the model will usually introduce enough phase error between $P_M$ and $P_W$ that the iterations will diverge. To handle this, one uses a zero phase low pass filter to cut off the learning above some frequency. This avoids divergence at the expense of not eliminating tracking error frequency components above the cutoff (Ref. 6). In the work reported here, we limit the investigation to model parameter error.

**THE SPEEDUP CONCEPT**

ILC can sometimes require many iterations to get to small error levels. This can be based on the choice of ILC law, some of which learn slowly at higher frequencies, or based on the limitation on the learning gain required for convergence, or based on the amount and frequency of model error, or based on the initial conditions created by the choice made for the initial run. Each iteration in ILC is an experimental run with hardware. Usually, our model of the world is reasonably good over many frequencies. This suggests that one could iterate with our model of the world, and perhaps reduce the tracking error substantially, and then convert to making experiments iterating with the hardware. Perhaps this could bypass many hardware iterations. For iterations using our model of the world we know everything we need in order to make such iterations, and they can be done numerically offline before starting the ILC hardware iterations. The concept is that the initial hardware run in this case, might start much closer to the converged input needed to reach zero tracking error, and thus substantially decrease the number of hardware iterations needed.

The procedure to use: Start with the initial condition hardware run, applying one's chosen initial input and observing the output and associated error $\underline{u}_0$ and $\underline{y}_0, \underline{e}_0$. Then update the error histories using one's model

$$\underline{e}_{M,j} = (I - P_M L_M)\underline{e}_{M,j-1} \tag{20}$$

For each iteration $j$, compute the associated ILC law update

$$\underline{u}_{M,j+1} = \underline{u}_{M,j} + L_M \underline{e}_{M,j} \tag{21}$$

It is this input history that is used to start the iterations in hardware. These iterations on the computer should be substantially simpler that running the iterations on hardware. In addition to that, the next section gives a way so simplify the computations, replacing square matrix multiplications by scalar multiplications. Going further, a formula is developed to find the result of such iterations without actually performing them.

Later we will discuss making the choice of how many iterations to make with the model before switching to hardware iterations. One choice is to try to achieve zero error in the model, but of course this control history does not produce zero error when applied to the world. Switching earlier could be beneficial. At the iteration of the switch, the $\underline{u}_{M,j}$ is applied to the world and can be renamed $\underline{u}_{W,0}$, The resulting output is used to start the hardware iterations, which we renamed $\underline{y}_{W,0}$, $\underline{e}_{W,0}$.

**FASTER COMPUTATIONS OF UPDATES WITH MODEL**

The main computation from iteration to iteration using the model, is to take $(I - P_M L_M)$ to higher and higher powers. The matrix dimension can be large, since it is based on the number of time steps in a run. It can be beneficial to diagonalize this matrix and define the error in a new error space (the Euclidian norm of the error is preserved)

$$(I - P_M L_M) = M_M \Lambda_M M_M^T \tag{22}$$

$$\Lambda_M = diag[\lambda_1 \quad \lambda_2 \quad \cdots \quad \lambda_N] \tag{23}$$

$$\underline{e}'_{M,j} = M_M \underline{e}_{M,j} \tag{24}$$

$$\underline{e}'_{M,j+1} = \Lambda_M \underline{e}'_{M,j} \tag{25}$$

$$\underline{u}_{M,j+1} = \underline{u}_{M,j} + L_M M_M^T \underline{e}'_{M,j} \tag{26}$$

Use Eq. (25) to update the error from one iteration to the next, then use Eq. (26) to know the ILC input update.

If you want to find the result after $N$ iterations, without bothering to find the previous error histories, one can find the sum of the powers of $\Lambda$'s directly for iteration $N$ using the following summation. Set

$$S = I + \Lambda + \Lambda^2 + \cdots + \Lambda^N$$

$$\Lambda S = \Lambda + \Lambda^2 + \Lambda^3 + \cdots + \Lambda^{N+1}$$

$$(I - \Lambda)S = I - \Lambda^{N+1}$$

$$S = (1 - \Lambda)^{-1}(I - \Lambda^{N+1}) \quad ; \quad \text{all } \lambda < 1 \tag{27}$$

Then the computation of the error history and the input history at iteration $N$ from the following formula bypassing any iterations. Given the initial hardware run that inputs whatever function one chooses, producing $\underline{u}_{M,0}, \underline{e}'_{M,0} = M_M \underline{e}_{M,0}, \underline{e}_{M,0} = \underline{y}^* - \underline{y}_0$, one uses result Eq. (27) to produce a simple formula that requires no iterations

$$\underline{u}_{M,N} = \underline{u}_{M,0} + L_M M_M^T [1 + \Lambda_M + \Lambda_M^2 + \cdots + \Lambda_M^N] \underline{e}'_{M,0}$$

$$\underline{u}_{M,N} = \underline{u}_{M,0} + L_M M_M^T (1 - \Lambda_M)^{-1} (I - \Lambda_M^{N+1}) \underline{e}'_{M,0} \tag{28}$$

$$\underline{e}'_{M,N} = \Lambda_M^N \underline{e}'_{M,0} = \Lambda_M^N M_M \underline{e}_{M,0} \tag{29}$$

## A CRITERION TO KNOW WHEN TO SWITCH

Investigate switching to hardware iterations after $N$ iterations using the model. We offer the following test to perform (on hardware) to help decide if you want to switch to hardware iterations.

(A1) At model iteration $N$ one knows $\underline{u}_{M,N}$ and $\underline{e}_{M,N}$. Compute the Root Mean Square of the error history

$$R_{M,N} = \sqrt{\frac{\underline{e}_{M,N}^T \underline{e}_{M,N}}{N}} \tag{30}$$

(A2) Perform one more iteration with the model

$$R_{M,N+1} = \sqrt{\frac{\underline{e}_{M,N+1}^T \underline{e}_{M,N+1}}{N}} \tag{31}$$

(B1) Apply input $\underline{u}_{M,N}$ to the hardware and observe the error history. Denote the RMS error as

$$R_{W,N} = \sqrt{\frac{\underline{e}_{W,N}^T \underline{e}_{W,N}}{N}} \tag{32}$$

(B2) Make one ILC iteration with the hardware, to produce RMS error

$$R_{W,N+1} = \sqrt{\frac{\underline{e}_{W,N+1}^T \underline{e}_{W,N+1}}{N}} \tag{33}$$

Then $R_{M,N} - R_{M,N+1}$ is the amount of error reduction in one model iteration. And $R_{W,N} - R_{W,N+1}$ it the amount of error reduction if one uses real world updates. Immediately after switching from model to world the RMS error should make a jump increase, because model error is now seen. The switch should be based on the slope of the error, pick the world hardware iterations if they learn substantially faster. Of course, ultimately the world is what matters.

## DEALING WITH UNSTABLE INVERSE PROBLEMS

Digital control systems usually have a plant described by a continuous time differential equation, and a digital controller whose input to the plant is through a zero order hold. The sampled output of the plant can be given by a difference equation that has the same solution as the differential equation at the sample times. For reasonable sample rates, independent of the number of zeros in the plant Laplace transfer function, the equivalent difference equation will introduce enough zeros to the plant $z$-transfer function so that there is one less zero than the number of poles. The result corresponds to a one time step delay between a change in input at a time step, and the first sample time step when a change in output is observed. In the case of a third order system with no zero in the Laplace transfer function, there will be two zeros in the $z$-transfer function. Reference (11) gives the locations when the sampling time interval becomes arbitrarily short. One zero is outside the unit circle on the negative real axis, the other is inside at the reciprocal location. A zero is introduced outside the unit circle for a plant with at least three more poles than zeros in its Laplace transfer function. When the closed loop $z$-transfer function is computed it will normally have the same zeros as the converted plant. The result is that the inverse problem produces an unstable input function, it has a pole or poles outside the unit circle. It is this unstable function that the ILC inverse problem is aiming to converge to as iterations progress. Many ILC users fail to notice this difficulty, the ILC seems to be mostly converged but at a disappointing error level. This very often happens, and is the result of the instability growing very slowly within the iterations.

A theory of "stable inverse" has developed to handle such situations (Ref. 19). The standard approach is somewhat award, and one of the current authors and co-workers have developed a number of alternatives (Ref. 13-16) to make ILC apply to such systems and result in a stable control input, is accomplished by not asking for zero error in the first time step of the desired trajectory if there is only one zero outside the unit circle, and not for two time steps if there are two zeros outside, etc. (Ref. 13). What is done for the first time step(s) is not really important as discussed in Ref. (20). The following modifications of the mathematics allow us to apply our speedup method to such problems. The input-output equation becomes

$$\underline{y}_D = P_D \underline{u} + \overline{A}_D x(0) \tag{34}$$

where the input $\underline{u}$ is full dimensional, and the first row or rows of $\underline{y}_D, P_D, \overline{A}_D$ as well as the desired output $\underline{y}_D^*$ are deleted corresponding to the number of initial time steps of error you want to ignore (greater than or equal to the number of zeros outside the unit circle). The definition of the error is similarly modified to $\underline{e}_D$ by initial row deletions. The learning control law becomes

$$\underline{u}_{j+1} = \underline{u}_j + L_D \underline{e}_{D,j} \tag{35}$$

The learning gain matrix $L_D$ becomes rectangular with fewer columns than rows. The solution to Eq. (32) for the needed input $\underline{u}^*$ to produce the desired output $\underline{y}_D^*$ is

$$\underline{u}^* = P_D^\dagger (\underline{y}_D^* - \overline{A}_D x(0)) \tag{36}$$

Here $P_D^\dagger$ is the Moore Penrose Pseudoinverse or other pseudoinverse. One can show that all singular values of the matrix are well behaved such that the needed input $\underline{u}^*$ does not exhibit unstable growth (Ref. 17). Of course, there are an infinite number of pseudo inverses including the true inverse without deleting. The true inverse that produces unstable input will normally be so large that no one would pick such an input. Hence, once can pick the initial inputs from the minimum Euclidean norm or Moore Penrose pseudo inverse to start the ILC, but any initial values that are far from the value needed by the true inverse also work. Reference (20) demonstrates the insensitivity to this choice. Moore Penrose minimizes what is done in the initial step(s). It is also best to pick a desired trajectory that starts up smoothly enough that the continuous time version of the desired trajectory does not require a Dirac delta function or unit doublet input to get onto the desired trajectory from the initial condition.

Taking the difference of Eq. (32) for run $j$ and run $j + 1$, and substituting the ILC law to eliminate the input from the equation, results in the error propagation equation

$$\underline{e}_{D,j} = (I - P_D L_D) \underline{e}_{D,j-1} \tag{37}$$

This error will converge to zero as $j \to \infty$ for all possible $\underline{e}_0$ if and only if the spectral radius of matrix $(I - P_D L_D)$ is less than unity, i.e. is $|\lambda_i(I - P_D L_D)| < 1$ for all eigenvalues. It will converge monotonically in the sense of the Euclidean norm of the error if the maximum singular value of this matrix is less than unity, $\sigma_i(I - P_D L_D) < 1$ for all $i$.

**NUMERICAL EXAMPLES**

We will study the speedup method applied to two systems, a second order system and a third order system. The first is described by

$$G(s) = \frac{\omega_n^2}{s^2 + 2\xi\omega_n + \omega_n^2} \tag{38}$$

with damping ratio $\xi$ and undamped natural frequency $\omega_n$. The model parameter values are $\xi_M = 0.5$ and $\omega_{Mn} = 37$ and the world values are $\xi_W = 0.3$ and $\omega_{Wn} = 37$. This transfer function is fed by a zero order hold sampling at 0.01 second. The chosen desired trajectory is $\underline{y}^*(t) = \pi(1 - \cos(20\pi t))^2$ for time steps from $1, 2, 3, \ldots, 100$. Each of the three ILC laws given in Eqs. (17)-(19) are investigated in the following figures, with parts (a), (b), and (c) of each figure applying to these three laws respectively.

Figure 2 illustrates issues about switching from iterations with the model, to iterations with the world. To create initial conditions for the iterations, one must apply an initial input history to the world, chosen here as $\underline{u}_0 = \underline{y}^*$, and record the output $\underline{y}_0 = P_M \underline{u}_0$ and the associated error $\underline{e}_0$. Curves of the RMS errors from initial run 0 to run 50, applies the ILC laws in Eqs. (17)-(19). The gain is $\phi = 1$ in all cases, which is small enough to make both the model iterations and the world iterations converge. The left of Fig. 2(a) shows the decay of the error from this starting condition when simply using our model to perform the 50 iterations. The dashed line shows what happens when one continues this beyond 50 iterations. Here one records the input history, the error, and the learning law input update, for each iteration. Then the RMS error for run 50 is computed, $R_{M,50}$, from Eq. (30) for item (A1) and the result is indicated on the Figure. One more iteration is performed with the model for item (A2). The value of $R_{M,51}$ is plotted, showing how much change in the model error is achieved in one iteration. This completes the black curve on subplots which are shown in the left column of Fig. 2.

To create the red curve, the input history obtained from model iteration at 50, is instead applied to the world and $R_{W,50}$ is computed from Eq. (32) for (B1). The result is shown on the plot, and it is a considerably larger tracking error than $R_{M,50}$ for (A1). This sudden increase in error is to be expected because (B1) contains the real world, and it is reacting to error that has not been seen by the iterations with the model. Then use an iteration in hardware to observe (B2) showing how much world error improvement has been obtained in one iteration.

The three different ILC laws investigated have different behaviors. The $P$ Transpose ILC is particularly robust, but at the expense of slow convergence for high frequencies. The learning rate is related to the square of the magnitude frequency response, which gets small at high frequencies. The range of learning gains that produce convergence when applied to a system with model error is enlarged. The Partial Isometry law learns quicker, its learning rate is related to the magnitude frequency response instead of its square. Normally this means that the convergence is faster. But if the gain used is close to the stability boundary for the actual error in the model, then convergence can become slow. The Quadratic Cost ILC law in most applications will behave similar to the *P* Transpose law.

To evaluate the benefit of the speedup, first realize that performing the iterations with the model is done by computer, offline, and does not require any experiments. Also important is that to examine what happens after a chosen number of iterations, it is possible to compute the result without doing any model iterations. One simply substitutes into the formulas in Eqs. (28)-(29). The implication is that instead of starting the iteration from the initial conditions generated as above, the formula finds the result of performing the 50 iterations from the initial condition. Then one can immediately start on the red curve. Thus, after using Eqs. (28)-(29), one immediately starts at (B1). Instead of an initial RMS error a bit above 15 dB, one starts the world hardware iterations at below 5 dB, and in addition, the error is decreasing fast. After 10 iterations, the RMS error is reduced to approximately -2 dB. One has bypassed performing about 60 iterations with the world hardware. Figure 3 displays the RMS history in a more representative way, making clear the benefit of the speedup. As explained, the speedup starts iterations with the world immediately achieving a much improved world error level.

The error in our model of the world, is perhaps a bit special. The world damping ratio is substantially less than that of the model. The world-model discrepancy is associated with a much increased resonant peak. We comment that the world iterations will decay to zero, but the model iterations result when applied to the world will approach a nonzero asymptote. Under the right conditions one might be able to start the world iterations below the converged model iterations. Using the parameters chosen for this numerical investigation we do not see the model iterations approaching a nonzero asymptote.

Deciding what model iteration number one should pick to make the conversion to world iterations is not obvious. It is suggested that if testing a candidate switch point, the world iterations have substantially faster decay in the comparison of (A1), (A2) to (B1), (B2), this could be a good indicator to switch. But this is only heuristic. The right hand side of Fig. 2(a) examines the benefit of using 100 model iterations versus 50 model iterations, before switching. For clarity the figure is a detail without showing the start of the model iterations. This change produces error below the model iteration curve. Of course, one can pick more than one candidate for the switch, and compute these four quantities. The only drawback is that these four quantities require iterations with the world.

Figure 3 presents this information in a way that represents what one does in practice and its benefits. The black curve gives the RMS error each iteration in the model updates. The curve can converge to zero with large enough repetition number, but of course this is zero error corresponding to our model, not zero error in the world. The blue curve gives the RMS error for world, applying the model results to the world. This curve when extended long enough will approach a non-zero positive error level.

The bottom curve (red curve) shows the benefit of using the speedup method to produce the starting point for the hardware ILC iterations. The red curve starts with the value from (B1) and starts the usual ILC hardware iterations. This makes the benefit of the speedup very obvious. The curve immediately has a much lower error level. Similar to Fig. 2, the subplots shown in the right column of Fig. 3 show the benefit of picking the start of the hardware iterations to begin after more model iterations. The number of model iterations is changed from 50 to 100.

Now consider the following third order system

$$G(s) = \left(\frac{a}{s+a}\right)\left(\frac{\omega_n^2}{s^2+2\xi\omega_n+\omega_n^2}\right) \tag{39}$$

with the model parameter values $a_M = 8.8, \xi_M = 0.5, \omega_{Mn} = 37$ and the world values $a_W = 8.8, \xi_W = 0.5, \omega_{Wn} = 44.4$. It is fed through a zero order hold. The chosen desired trajectory is $\underline{y^*(t)} = \pi(1 - \cos(10\pi t))^2$ for time steps from $2, 3, 4, \ldots, 100$, with the sample time interval chosen as 0.01 second. Note that for the third order system the pole excess is 3, which gives the sampled system one zero outside the unit circle provided the sampling period is sufficiently small. To address this, we don't ask for zero error in the first time-step of the desired trajectory. The first row of $P$ and first column of $L$ are deleted producing $P_D$ and $L_D$. Following Eqs. (17)-(19), we formulate the deleted ILC laws as

$$L_{D,M} = \phi_D P_{D,M}^T \tag{40}$$

$$L_{D,M} = \phi_D (V_M U_M^T)_D : \quad P_M = U_M \Sigma_M V_M^T \tag{41}$$

$$L_{D,M} = (\phi_D I + P_{D,M}^T P_{D,M})^{-1} P_{D,M}^T \tag{42}$$

Figures 4 and 5 are analogous to Figs. 2 and 3. The gain is $\phi_D = 1$ in all cases, which is small enough to make both the model iterations and the world iterations converge. Figure 4 illustrates issues about switching from iterations with the model, to iterations with the world. Figure 4 shows the decay of the RMS world error during iterations updating the input based on the model. Again, the dashed curve will approach a nonzero value so that the red curve is guaranteed to cross below the dashed black curve after a sufficient number of iterations.

On the left of Fig. 5(a) we see the benefit of picking the switch to world hardware iterations from the result associated with 50 model iterations, and on the right gives the result starting from 100 iterations. Each case starts from the desired trajectory as the initial run. The formula for the result at 50 or 100 iterations can be obtained simply by plugging into the formula derived, and using this as the starting point for ILC iterations in hardware produces the red curves in Fig. 5. Simply using this precomputed starting point, allow the hardware iterations to start at an RMS error approximately at 4 dB instead of 15 dB. Using 100 in place of 50 model iterations as the starting point, Fig. 5(a) left vs. Fig 5(a) right produces the hardware decay Red Curve reach lower error levels as the iterations progress.

For comparison, Fig. 5 include a blue curve and a black curve which show what happens if one simply started doing hardware ILC iterations from the beginning, and also the black curve which shows what the iterations with the model do starting from the same point. The benefit of the speedup method is very obvious.

## CONCLUSIONS

It can often happen that ILC learns an input producing very low error levels in rather few iterations. But this property is dependent on many different factors. The learning rate can be slow for errors at high frequencies when the response level is small. The learning gain must be picked so that the not only do iterations with the model converge, but iterations with the world must also converge. If the unknown model error is such that the model of the world differs from the model in such a way that it is close to the stability boundary, then convergence is slow. It may also be that hardware and sensors are very accurate, and one simply wants to converge to the lowest possible error level, by taking a large number of iterations. But the speedup approach presented here, one might decide to routinely use, simply because it can be so easy and effective.

Figures 3 and 5 are figures that graphically demonstrate the benefit of the speedup approach. It can be done quickly, and have very reduced RMS error levels immediately when one starts iterations in hardware, which can bypass many hardware runs. By simply using the formulas derived to obtain the result at any chosen model iteration, one takes advantage of immediately learning how to correct error in one's model, without having to perform ILC hardware iterations. And then one switches to ILC repetitions in hardware to make the corrections needed to get zero error in the world.

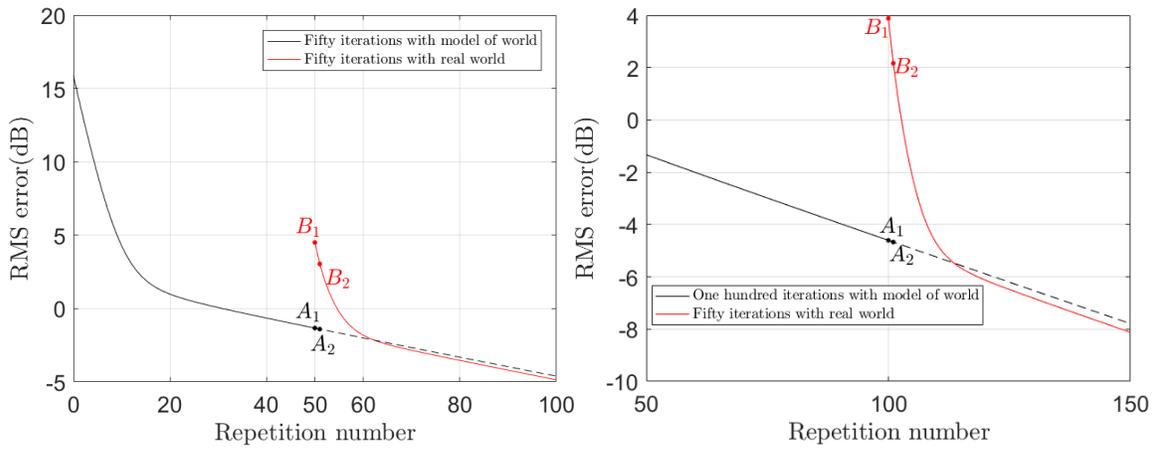

Figure 2(a). Illustration of the switching decision for the second order system using $L_M = \phi P_M^T$. The RMS error at $A_1, B_1$ is computed as $R_{M,50}, R_{W,50}$ (left) or $R_{M,100}, R_{W,100}$ (right)

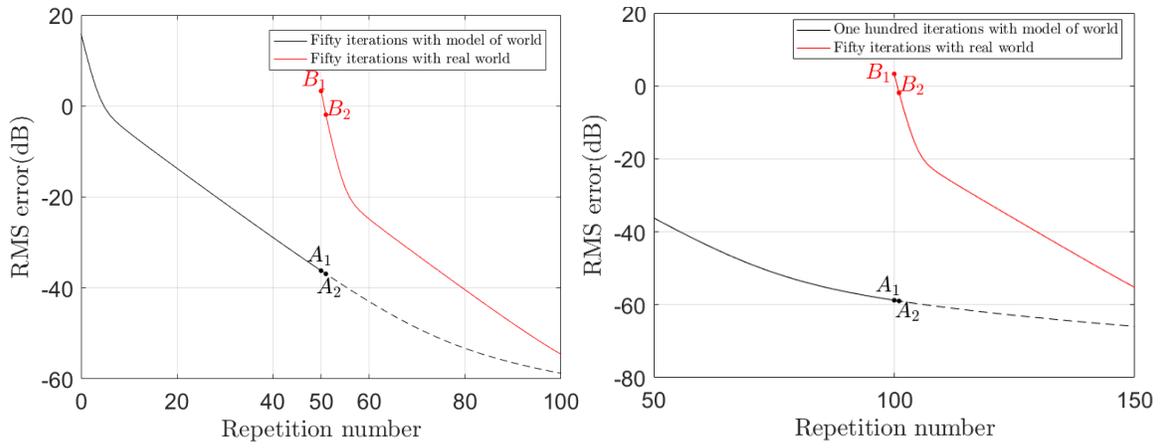

Figure 2(b). Illustration of the switching decision for the second order system using $L_M = \phi V_M U_M^T$. The RMS error at $A_1, B_1$ is computed as $R_{M,50}, R_{W,50}$ (left) or $R_{M,100}, R_{W,100}$ (right)

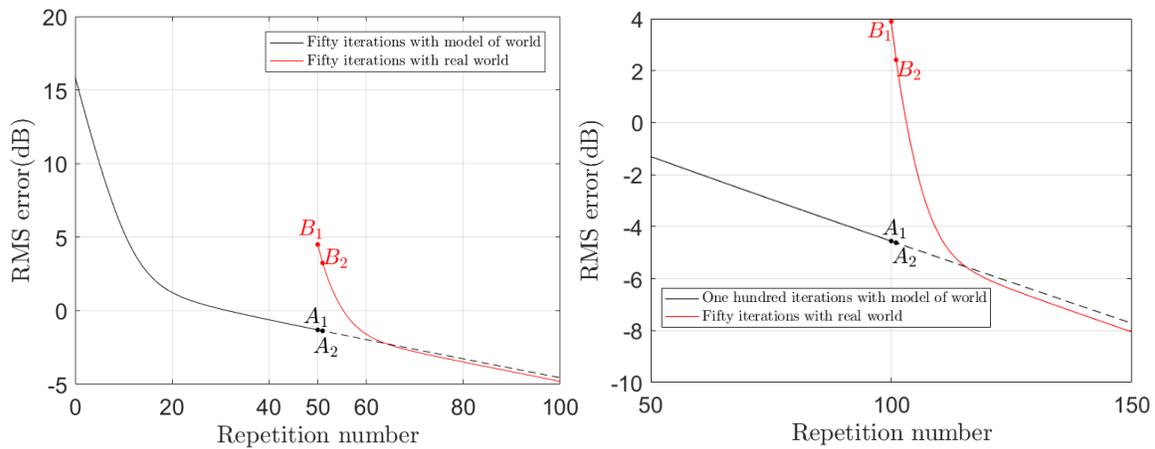

Figure 2(c). Illustration of the switching decision for the second order system using $L_M = (\phi I + P_M^T P_M)^{-1} P_M^T$. The RMS error at $A_1, B_1$ is computed as $R_{M,50}, R_{W,50}$ (left) or $R_{M,100}, R_{W,100}$ (right)

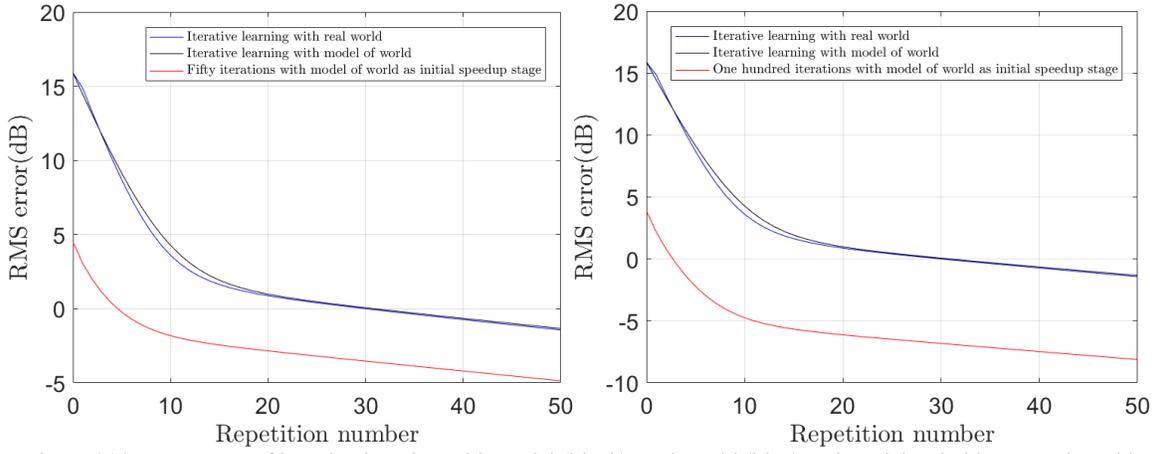

Figure 3(a). RMS error of iterative learning with model (black), real world (blue) and model switching to real world (red) for the second order system. The ILC law is $L_M = \phi P_M^T$ and the switching starts at 50 (left) iteration with model or 100 (right)

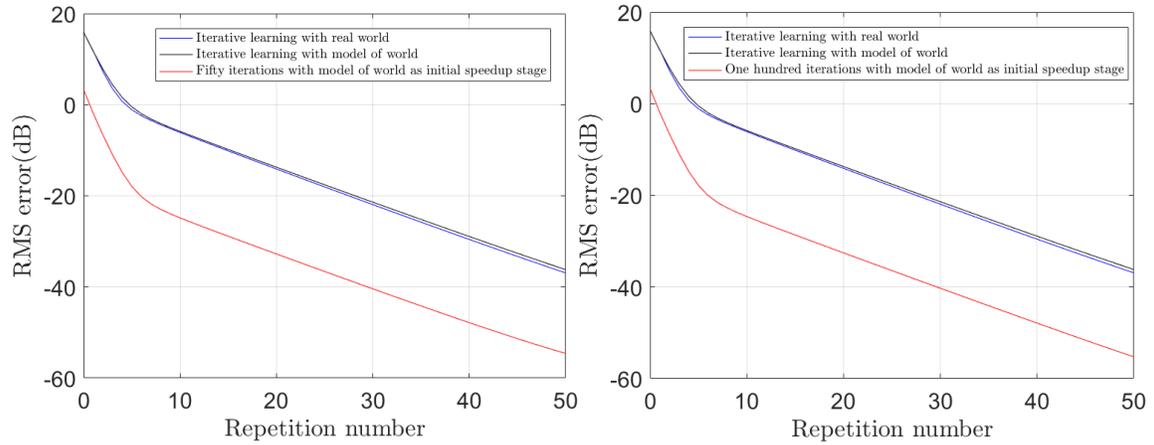

Figure 3(b). RMS error of iterative learning with model (black), real world (blue) and model switching to real world (red) for the second order system. The ILC law is $L_M = \phi V_M U_M^T$ and the switching starts at 50 (left) iteration with model or 100 (right)

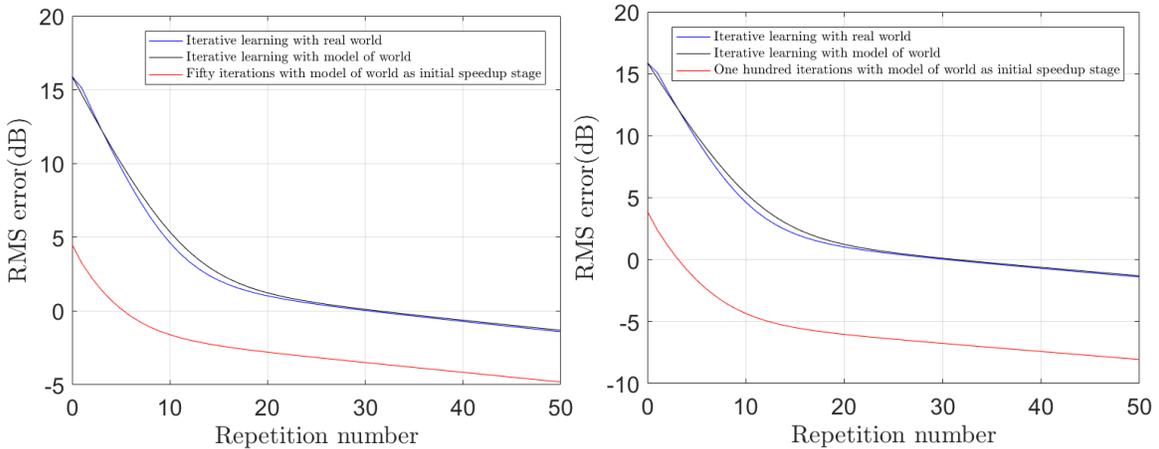

Figure 3(c). RMS error of iterative learning with model (black), real world (blue) and model switching to real world (red) for the second order system. The ILC law is $L_M = (\phi I + P_M^T P_M)^{-1} P_M^T$ and the switching starts at 50 (left) iteration with model or 100 (right)

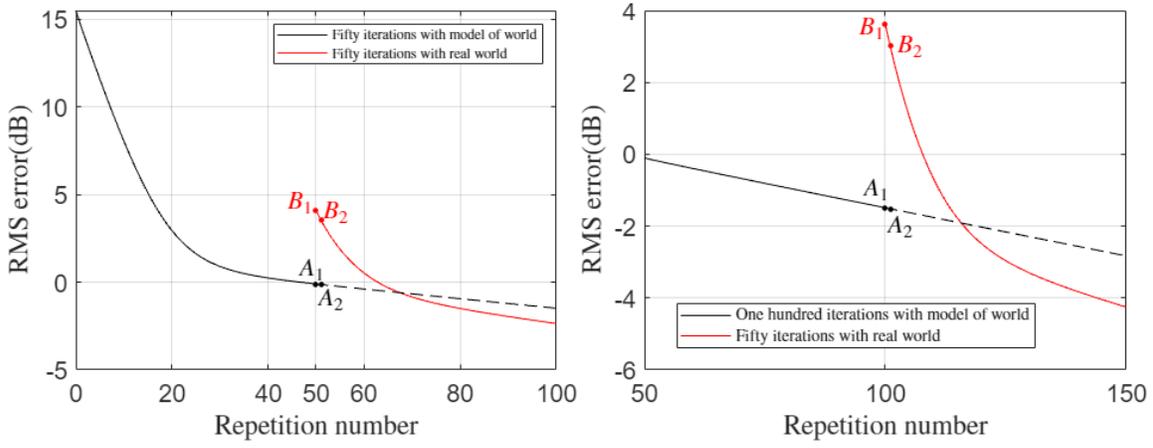

Figure 4(a). Illustration of the switching decision for the third order system using $L_{D,M} = \phi_D P_{D,M}^T$. The RMS error at $A_1, B_1$ is computed as $R_{M,50}, R_{W,50}$ (left) or $R_{M,100}, R_{W,100}$ (right)

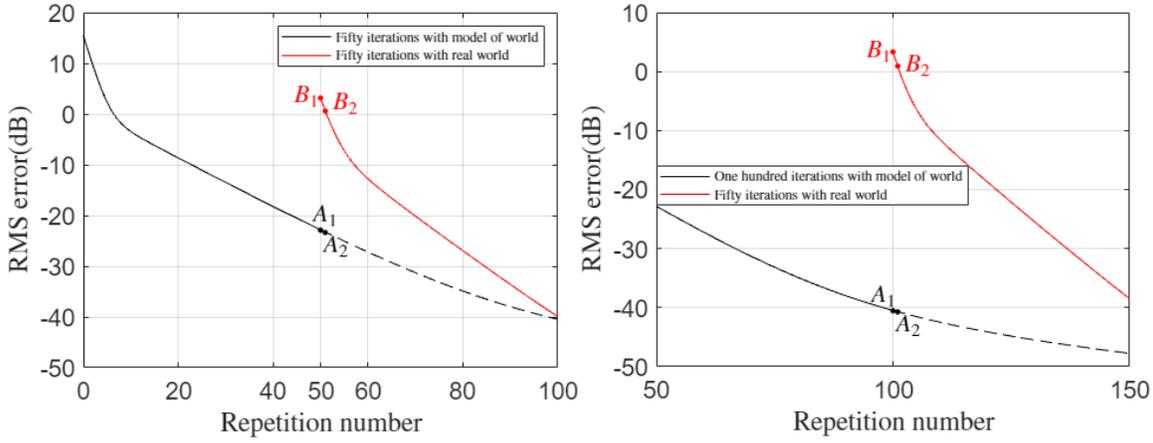

Figure 4(b). Illustration of the switching decision for the third order system using $L_{D,M} = \phi_D (V_M U_M^T)_D$. The RMS error at $A_1, B_1$ is computed as $R_{M,50}, R_{W,50}$ (left) or $R_{M,100}, R_{W,100}$ (right)

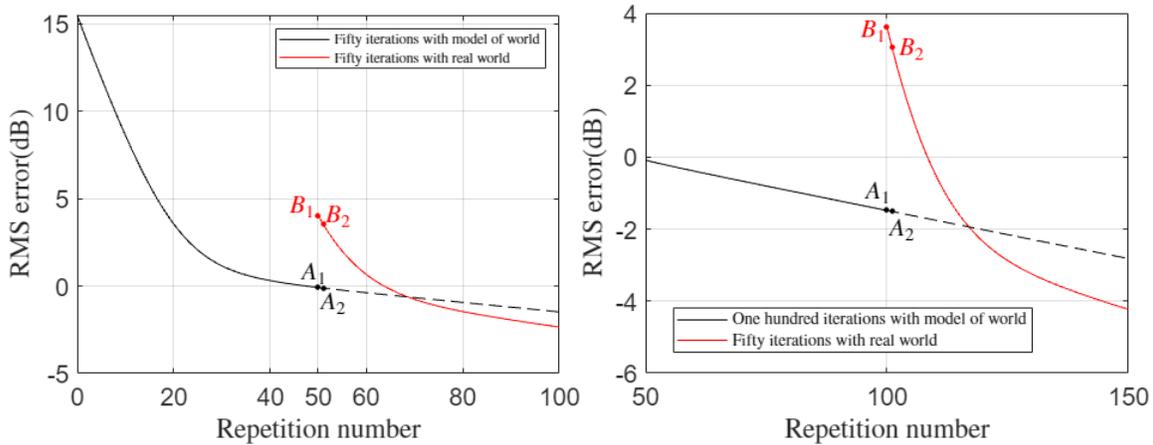

Figure 4(c). Illustration of the switching decision for the third order system using $L_{D,M} = (\phi_D I + P_{D,M}^T P_{D,M})^{-1} P_{D,M}^T$. The RMS error at $A_1, B_1$ is computed as $R_{M,50}, R_{W,50}$ (left) or $R_{M,100}, R_{W,100}$ (right)

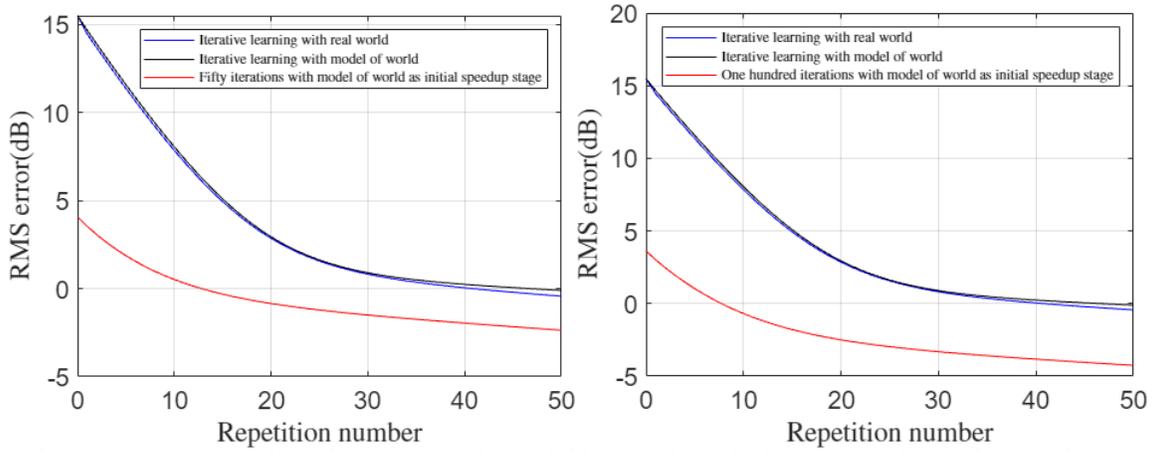

Figure 5(a). RMS error of iterative learning with model (black), real world (blue) and model switching to real world (red) for the third order system. The ILC law is $L_{D,M} = \phi_D P_{D,M}^T$ and the switching starts at 50 (left) iteration with model or 100 (right)

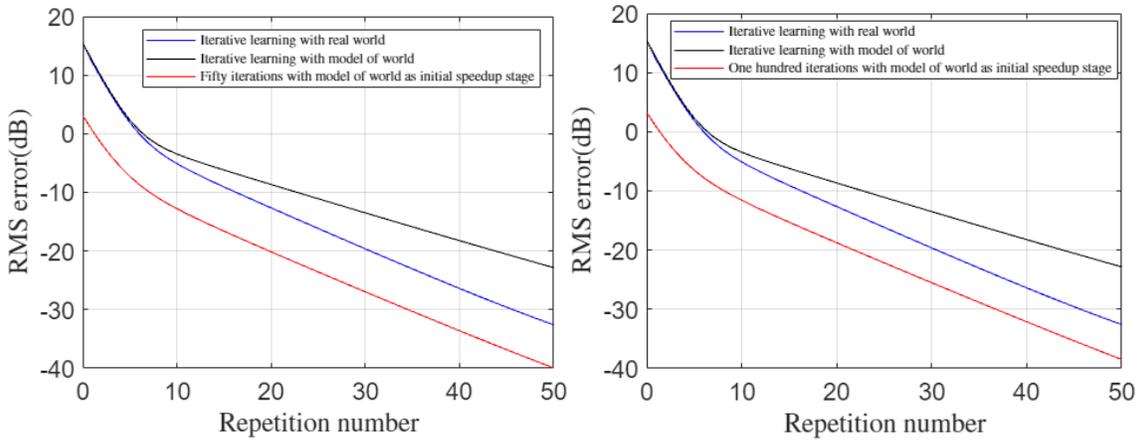

Figure 5(b). RMS error of iterative learning with model (black), real world (blue) and model switching to real world (red) for the third order system. The ILC law is $L_{D,M} = \phi_D (V_M U_M^T)_D$ and the switching starts at 50 (left) iteration with model or 100 (right)

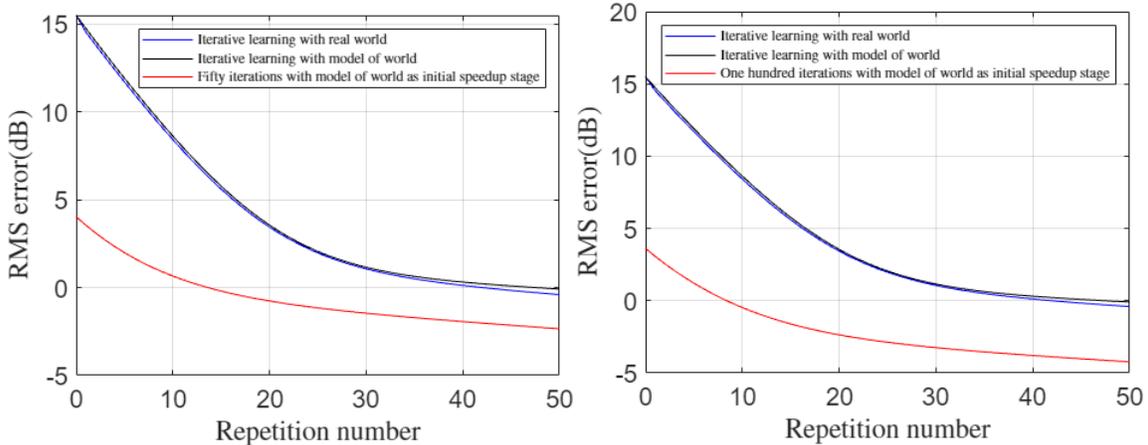

Figure 5(c). RMS error of iterative learning with model (black), real world (blue) and model switching to real world (red) for the third order system. The ILC law is $L_{D,M} = (\phi_D I + P_{D,M}^T P_{D,M})^{-1} P_{D,M}^T$ and the switching starts at 50 (left) iteration with model or 100 (right)